\documentstyle[twocolumn,prl,aps]{revtex}
\begin{document}
\twocolumn[
\hsize\textwidth\columnwidth\hsize\csname@twocolumnfalse\endcsname
\title{
      The nature of the short wavelength excitations in vitreous silica: \\
       X-Rays Brillouin scattering study.
      }
\author{
        O.~Pilla$^1$,
        A.~Cunsolo$^2$,
        A.~Fontana$^1$,
        C.~Masciovecchio$^3$,
        G.~Monaco$^4$, \\
        M.~Montagna$^1$,
        G.~Ruocco$^5$,
        T.~Scopigno$^1$, and
        F.~Sette$^4$.
        }
\address{
         $^{1}$ Dipartimento di Fisica and INFM,
                Universit\`a di Trento, I-38050 Povo, Trento, Italy\\
         $^{2}$ Dipartimento di Fisica and INFM,
                Universit\`a di Roma III, I-00146 Roma, Italy\\
         $^{3}$ Sincrotrone Trieste, Area Science Park,
                I-34012, Trieste, Italy. \\
         $^{4}$ European Synchrotron Radiation Facility,
                BP 220, F-38043 Grenoble Cedex, France \\
         $^{5}$ Dipartimento di Fisica and INFM,
                Universit\`a dell'Aquila, I-67100, L'Aquila, Italy
         }
\date{\today}
\maketitle
\begin{abstract}

The dynamical structure factor ($S(Q,E)$) of vitreous silica has
been measured by Inelastic X-ray Scattering varying the exchanged
wavevector ($Q$) at fixed exchanged energy ($E$) - an experimental
procedure that, contrary to the usual one at constant $Q$, provides
spectra with much better identified inelastic features. This allows 
the first direct evidence of Brillouin peaks in the $S(Q,E)$ of 
$SiO_2$ at energies above the Boson Peak (BP) energy, a finding 
that excludes the possibility that the BP marks the transition 
from propagating to localised dynamics in glasses.
\end{abstract}
]

The nature of the high frequency collective vibrations in glasses
is highly debated in the physics of
disordered materials \cite{gen}. At long wavelengths, these
vibrations are thought to be well approximated by propagating
plane waves (sound waves) with a linear relationship between 
energy, $E$, and wavevector, $Q$. Increasing $Q$, and in
particular when $Q$ becomes comparable to the inverse of the
interparticle separation $a$, the simple linear relation between
$E$ and $Q$ no longer holds, and the dispersion relation shows a
bend down, a maximum at $Q \approx \pi/a$ and a broad minimum at
$Q \approx 2\pi/a$: this behaviour closely resembles that of the
phonon dispersion relation in crystals. At variance with this
apparent crystal-like behaviour, several evidences indicate that,
even in the low $Q$ limit, the nature of the excitations in glasses
is by far more complicated than in crystals. Among these evidences
there are the universal low temperature anomalies observed in the 
heat capacity and heat conductivity \cite{ph,zp,phill}, and the
presence of the so called Boson Peak in Raman \cite{fonta,soko}
and Inelastic Neutron Scattering (INS)\cite{buchen} spectra.
Despite the experimental and theoretical efforts that in the last
years have been devoted to the characterisation of the excitations
in disordered systems, the central question about this issue,
namely the propagating or localised character of the collective
excitations, is not completely settled.
\cite{sio2noi,andalo,sio2loro,duval,aleorba}

Two main interpretations of the existing experimental data have
been proposed so far: they lead to a contrasting view on the nature
of the collective excitations in the mesoscopic $Q$-$E$ region
($Q$=1-10 nm$^{-1}$) \cite{gen}. One view assumes that the static 
disorder in the glass produces strong elastic scattering of sound 
waves and, as a consequence, the vibrational modes at become 
localised above a cross-over energy ($E_{CO}$). In particular, 
it is assumed that the Boson Peak energy corresponds to this 
cross-over energy marking the onset of localisation. The second 
point of view is based on experimental evidences indicating that 
the vibrations in glasses maintain a collective propagating nature
even at energies well above the Boson peak energy. In this respect, 
the high energy dynamic structure factor ($S(Q,E)$) of different 
glasses and glass-forming liquids, ranging from fragile systems 
to strong ones, have been recently investigated by the new 
Inelastic X-ray Scattering (IXS) technique \cite{GlassButSiO2}. 
These studies have shown the existence of inelastic features 
whose energy $\Omega_Q$ has an almost linear dispersion with $Q$. 
This extends up to $Q$ values few tenth of $Q_o$ (the position 
of the first maximum in the static structure factor $S(Q)$) 
and up to energies in the Boson Peak energy region. These findings
demonstrate that propagating collective excitations persist in the 
mesoscopic $Q$ region, and, in particular, that the Boson Peak 
energy ($E_{BP}$) cannot be straightforwardly related to the 
cross over energy marking the transition between propagating 
and localised vibrations.

The IXS data reported so far for vitreous silica (v-$SiO_2$),
despite the generality of the IXS data on the other studied
glasses which support the second point of view, are utilised 
to support either the localised \cite{sio2loro}, or the 
propagating \cite{sio2noi,andalo} picture. This conflicting 
situation is particularly unsatisfactory considering that 
v-$SiO_2$ is the prototype of strong glasses. This can be partly 
ascribed to the low intensity of the inelastic signal compared 
to the tail of the resolution broadened elastic one. Indeed, 
even at the highest investigated temperature, the inelastic 
contribution in the IXS spectra of v-$SiO_2$ is observed as 
a weak shoulder of the elastic line as can be seen in Fig.~2 of
Ref.~\cite{andalo}. As a consequence, in order to deduce from
The IXS data the existence of Brillouin peaks and compare their
lineshapes to theoretical predictions, the data must be fitted to
a theoretical model and deconvoluted from the instrumental
response function. As a matter of fact, these experimental
data do not give a clear-cut indication on the choice among
contrasting theoretical models. Even in a recent work on
densified $SiO_{2}$ glass \cite{ericden}, where the inelastic peak
was slightly better resolved than in v-$SiO_2$, the Brillouin
intensity was so weak that a considerable effort of subjective data
processing was needed to compare the experimental result to
theories.

In this paper we present a {\it direct experimental evidence} on
the existence of well defined Brillouin peaks in the IXS spectra
of amorphous silica. At variance with the previous IXS experiments 
on v-$SiO_2$ \cite{sio2noi,andalo,sio2loro,ericden}, where the 
spectra were measured at constant $Q$ and as a function of $E$, 
here we present constant $E$ cuts of $S(Q,E)$ taken as a function 
of $Q$. To appreciate the advantage of this approach, we remind 
that in the constant $Q$ spectra, as a consequence of the finite 
energy resolution, the elastic component gives rise to a 
$\sim E^{-2}$ tail whose intensity in the Brillouin peak region 
is often stronger than that of that of the peak itself. 
Therefore, the extraction of the spectroscopic parameters of 
the Brillouin peak necessarily requires a fitting procedure 
and the assumption of a model function for the $S(Q,E)$. On 
the contrary, in the constant $E$ measurements -the procedure 
utilised in the present experiment- the elastic contribution,
convoluted to the instrumental response function, gives rise to an
almost $Q$-independent background which should not affect
significantly the position and lineshape of the inelastic signal.
This approach allows us to demonstrate the presence of a 
Brillouin peak at an energy larger than the $E_{BP}$. 
Therefore, without any specific model for the $S(Q,E)$, we
demonstrate that the Boson Peak does not mark the transition between
propagating and non-propagating dynamics in vitreous silica.

The experiment has been carried out at the very high energy
resolution IXS beamline (ID16) at the European Synchrotron
Radiation Facility. The instrument consists of a back-scattering
monochromator and five independent analyser systems, held one next
to each other with a constant angular offset on a 7~m long
analyser arm. We utilised the Si(11~11~11) configuration, giving a
total instrumental energy resolution of 1.5~meV
full-width-half-maximum (fwhm)\cite{mascio}, and an offset of 3
nm$^{-1}$ between two neighbour analysers. The momentum transfer,
$Q=2k_{\circ}sin(\theta_s/2)$, with $k_{\circ}$ the wavevector of
the incident photons and $\theta_s$ the scattering angle, is
selected by rotating the analyser arm. Few spectra at constant $Q$
and as a function of $E$ ($Q$=20, 23, 26, 29 and 32 nm$^{-1}$)
were measured with a $Q$ resolution of 0.4 nm$^{-1}$ fwhm for
normalisation purposes. The spectra at constant $E$ and as a
function of $Q$ ($E^*$=0, 5.3 and 8.5 meV) were made in two steps.
In a first measurement, the spectra were taken in the -2$\div$32
nm$^{-1}$) range by using the five analysers. In a further
measurements the -2$\div$6 nm$^{-1}$ region were studied in more
detail and with a better accuracy using the analyser number two
with a $Q$ resolution of 0.4 nm$^{-1}$ fwhm. The obtained
individual spectra have been then pasted together. The $E$ scans
at constant $Q$ were performed by varying the monochromator
temperature with respect to that of the analyser crystals. The
$Q$-scans at constant $E$ were performed by setting a constant
temperature offset between the monochromator and the analyser
crystals and by rotating the analyser-arm to vary the scattering
angle. Each spectrum, either at constant $Q$ or $E$, took about
150', and each fixed-$Q$ or fixed-$E$ point was obtained by
typical average of five scans. The data were normalised to the
intensity of the incident beam. The sample, a Spectrosil-grade
$SiO_2$, was placed in a variable temperature oven stabilised at
1200 K within $\pm$ $10$ K. This high temperature was used 
to enhance the inelastic signal. 

The spectra taken at constant energy $E^*$=0, 5.3 and 8.5 meV
are reported in Figs.~1a and b. For clarity the error bars are not
reported, these are of the order of the scattering of the data
with respects to a smooth behaviour. Obviously, increasing energy,
the scattering intensity decreases and the spectra become more
noisy. The two energy transfers were chosen to be such to  
correspond to $E_{BP}$ ($E^*$=5.3 meV) and on the high frequency 
side of the Boson Peak ($E^*$=8.5 meV). The $E^*$=0 spectrum 
has been collected in order to establish the elastic contribution
to the spectra at finite energy transfer.
This latter spectrum shows, in the
measured $Q$ range, the intense feature usually
referred to as the First Sharp Diffraction Peak (FSDP) centred at
about 15 nm$^{-1}$ and, more importantly, it is 
almost flat in the 2-6 nm$^{-1}$ $Q$-region, where one should
expect Brillouin peaks.

Directly in the raw data of Fig.~1 one can see at low $Q$ the
existence of well defined structures peaking at 1.25 nm$^{-1}$ in
the $E^*$=5.3 meV cut and 2.1 nm$^{-1}$ in the $E^*$=8.5 meV cut.
These two $Q$ positions change almost proportionally to $E^*$. The
peaks are superimposed to the featureless elastic contribution. 
The presence of these peaks in the constant energy spectra
is the fingerprint for the existence of vibrational excitations
bearing a strict relationship between energy and wavevector.
This, in turns, can be seen as a demonstration of the
spatially non-localised nature of these excitations, and, in
particular, of their propagating character.

The spectra in Fig.~1 contains an elastic contribution as a consequence
of the finite energy resolution. In order to subtract the elastic
part a series of spectra at constant $Q$ have been collected.
Two examples of these spectra are reported in Fig.~2a and b for
the constant $Q$ cuts of $Q$=23 and 32 nm$^{-1}$  (circles). 
The resolution function (full line), aligned with the elastic peaks,
is also reported to emphasise the presence a clear inelastic 
contribution even at these high  $Q$ values. These data have 
been used to normalise the relative intensity between the
inelastic and elastic spectra shown in Fig.~1 according to the 
following procedure. After aligning and scaling the experimentally 
determined resolution function to the elastic peaks in the spectra 
of Fig.~2, we estimate the relative intensity between the elastic 
and inelastic signals at the energy transfers, $E^*$, utilised in 
the constant-$E$ cuts spectra (arrows in Fig.~2a). This
elastic to inelastic intensity ratio have been obtained at $Q$=20,
23, 26, 29 and 32 nm$^{-1}$. These ratios allows then to put 
in the correct relative scale the spectra taken at $E^*$$\neq$$0$ 
and at $E^*$=0. The consistency of this procedure can be
appreciated in Fig.~1 considering that the elastic 
contribution is shown on a scale determined by the ratio measured 
at $Q$=23 nm$^{-1}$ in the spectra in Fig.~2a, and that the 
crossed circles in Fig.~1b are the calculated values of total 
(elastic plus inelastic) scattering derived from the other constant 
$Q$ scans. 

This normalisation procedure is used to derive the inelastic 
part of the $S(Q,E)$ by the subtraction of the normalised 
elastic contribution from the total scattering intensity. 
The difference spectra are reported in Fig.~3 (circles) 
together with the error bars as derived from the counting 
statistics. In these spectra the existence of well defined
Brillouin peaks is highly emphasised. Let us now compare 
the inelastic spectra reported in Fig.~3 with the predictions 
of the two models proposed in the literature as a "good" 
representation of the dynamics structure factor of vitreous 
silica. In the first one, the $S(Q,E)$ is derived within the 
framework of the generalised Langevin equation for the considered 
$Q$-component of the density fluctuation correlator \cite{balucani}, 
with a choice of the memory function, $m_Q(t)=2 \Gamma_Q
\delta(t) + \Delta_Q^2$ \cite{giulio}. This choice is appropriate 
for glassy systems. The time independent term
$\Delta_Q$, whose value determines the change of the sound
velocity, $c$, between the fully relaxed ($c_o$) and unrelaxed
($c_\infty$) limiting values, reflects the presence of the frozen
structural $\alpha$-relaxation. The parameter $\Gamma_Q$
determines the width of the side peaks, i.~e. the sound wave
attenuation coefficient, and summarise the effects of very fast
(or "instantaneous") sound absorption mechanisms. In
Ref.s~\cite{NonDyn,relH} these mechanisms have been ultimately
associated with the presence of structural disorder in the glass.
Using the previous expression for $m_Q(t)$, the $S(Q,E)$ turns out
to be:
\begin{equation}\label{dho}
\frac{S(Q,E)}{S(Q)}=f_Q \; \delta(E) + \frac{1\!-\!\!f_Q}{\pi}
\frac{\Omega_Q^2 \Gamma_Q}{(E^2-\Omega_Q^2)^2+E^2 \Gamma_Q^2},
\end{equation}
i.~e. the sum of an elastic line of intensity $f_Q$ ($f_Q$ is the
non-ergodicity parameter, $f_Q=1-(c_o/c_\infty)^2$), and of an
inelastic contribution, whose shape is known as Damped Harmonic
Oscillator (DHO).
In previous IXS studies on v-$SiO_2$, Eq.~1 was used to fit the
constant-$Q$ measurements leaving $\Gamma_Q$ and $\Omega_Q$ as
free parameters. In these studies turned out that, in the small
$Q$ limit, $\Omega_Q$=$vQ$ ($v$=6800 m/s) \cite{andalo} 
and $\Gamma_Q$=$DQ^2$ ($D=1.3$ meV/nm{-2}) \cite{sio2noi}.

Differently from the Langevin equation approach, the authors of
Ref.~\cite{sio2loro} proposed a model (the so-called EMA model)
for the $S(Q,E)$ where its inelastic part is assumed to be:

\begin{equation}\label{ema}
S(Q,E)\propto \frac{c_E^2Q^4}{E}
\frac{\Gamma_E}{((E^2+\Gamma_E^2-c_E^2Q^2)^2+4\Gamma_E^2c_E^2Q^2)}
\end{equation}
where $\Gamma_E$ and $c_E$ are given by:
\begin{eqnarray}\label{emagamma}
\Gamma_E={E^4}/{E_{CO}^3}\;[1+({E}/{E_{CO}})^m]^{-3/m}
\\ \label{emac}
 c_E=c_0 \; [1+({E}/{E_{CO}})^m]^{z/m}
\end{eqnarray}
Fitting the IXS data of Ref.~\cite{sio2noi}, the authors of
Ref.~\cite{sio2loro} found for the free parameters entering in
Eqs.~\ref{ema}-\ref{emac} the values: $m$=2, $E_{CO}$=3.9 meV, 
$c_0$=5900 m/s, and $z$=0.37.

In Fig.~3 we report as solid and dashed lines the inelastic part of
the $S(Q,E)$ obtained by Eqs.~{\ref{dho}} and {\ref{ema}}
respectively convoluted with the experimental resolution function.
In calculating the theoretical expressions for the $S(Q,E)$ we
have used the values of the parameters listed before. The only adjustable
parameter is an intensity factors, that has been arbitrarily
chosen in order to best fit the experimental data. As can be seen, 
the DHO model is in a good agreement with both of the constant-$E$ 
spectra in explaining the presence of the Brillouin peak and the 
high $Q$ plateau, although the relative intensity of these two 
main features is not completely accounted for. On the contrary
the EMA model fails to reproduce the existence of the observed 
Brillouin peak. This is not surprising as the EMA models is based 
on the assumption that at $\approx$5 meV (the BP energy) the 
vibrational modes in vitreous silica became localised. This 
assumption can be now definitively discarded by the direct 
inspection of the raw data.

The data $E^*$=5.3 meV clearly discriminate between DHO and EMA
models, showing a striking agreement with the DHO. Also the data
at $E^*$=8.5 favour the DHO model predictions, although the higher 
noise level precludes a conclusion as firm as in the $E^*$=5.3 meV
case. 

In conclusion, this study, independently of any specific model of
the $S(Q,E)$, has experimentally demonstrated that in v-$SiO_2$
there are collective propagating excitations at energies above the
Boson peak energy. Moreover, it has shown that the model of the
$S(Q,E)$ based on the generalised Langevin equation approach is
successful in reproducing the main features observed in the 
experimental data.

{\footnotesize{

\begin{description}

\item  {
Fig.1 - 
IXS spectra
of amorphous silica at T = 1200 K taken as function of $Q$ at
constant $E$: (a) $E^*$=8.5 meV (circles) and $E^*$=0 meV (full line); 
(b) $E^*$=5.3 meV (circles) and $E^*$=0 meV (full line). The arrows
indicate the $Q$ positions of the Brillouin peaks. The crossed
circles are the total scattered intensity derived from the
constant $Q$ cuts at energies of $E^*$=8.5 and 5.3 meV.}

\item  {
Fig.2 - 
IXS spectra
of v-$SiO_2$ at T = 1200 K taken at constant $Q$ (circles): (a)
$Q$=23 nm$^{-1}$ and (b) $Q$=32 nm$^{-1}$. The full lines are the
resolution function aligned with the elastic peaks. The arrows
indicate the energies values of 5.3 and 8.5 meV where the constant-$E$ 
cuts were measured.
}

\item  {
Fig.3 - 
The inelastic
contribution to the spectra reported in Fig.~1 (circles) are
reported together with their error bars in an expanded $Q$ scale:
(a) $E^*$=8.5 meV; and (b) $E^*$=5.3 meV. The full (dashed) lines are the
parameter-free lineshapes calculated via Eq.~\ref{dho} (Eq.~\ref{ema}).
The two theoretical predictions have been multiplied by arbitrary scale
factors.
}
\end{description}
}}

\end{document}